\documentclass[10pt,draft]{dis03}
\usepackage{epsf,amsmath}

\newcommand{\be}{\begin{equation}}
\newcommand{\ee}{\end{equation}}
\newcommand{\bea}{\begin{eqnarray}}
\newcommand{\eea}{\end{eqnarray}}
\newcommand{\ba}{\begin{array}}
\newcommand{\ea}{\end{array}}
\newcommand{\bean}{\begin{eqnarray*}}
\newcommand{\eean}{\end{eqnarray*}}
\newcommand{\non}{\nonumber}

\newcommand{\bc}{\begin{center}}
\newcommand{\ec}{\end{center}}

\begin{document}
\title{
\begin{flushleft}
{\sf SFB/CPP-03-29}
\end{flushleft}
NLL QED CORRECTIONS TO DEEP INELASTIC SCATTERING 
\thanks{This project was supported in part by SFB-TR 09.}}

\author{Johannes Bl\"umlein and Hiroyuki Kawamura 
\footnote{The present address is KEK, OHO 1-1, Tsukuba, 301-0801
Ibaraki, Japan}\\
Deutsches Elektronen Synchrotron, DESY\\ 
Platanenallee 6, D--15738 Zeuthen, Germany}

\maketitle

\vspace*{2mm}
\begin{abstract}
\noindent 
The $O(\alpha^2\log(Q^2/m_e^2))$ leptonic QED corrections to unpolarized 
deeply inelastic electron-nucleon scattering are calculated in the mixed 
variables. 
\end{abstract}

\section{Introduction} 
Deep inelastic scattering provides us with detailed information on the 
nucleon structure. In order to extract the parton distribution functions 
from DIS cross sections and to measure $\alpha_s(M_Z^2)$ with high 
precision it is crucial to 
control the QED  radiative corrections. 
The 1-- and 2--loop leading--log QED corrections 
were derived in Ref.~[1--3]. Complete 1-loop corrections  for DIS were 
given in Refs.~\cite{ABKR}. Furthermore the universal leading 
logarithmic corrections were derived to $O((\alpha L)^5)$ both 
for 
polarized and unpolarized processes in 
\cite{JBHK1}, where also the resummation of the $O((\alpha \ln^2(z))^k)$
for polarized scattering was given. In this paper, we summarize our recent 
results of NLO leptonic 
QED corrections in mixed variables \cite{JBHK2}.

\vspace*{-1.5mm}
\section{Mixed variables}
In general radiative corrections do strongly depend on how the kinematic 
variables are measured. In this paper, we consider  the case of mixed 
variables, i.e. $y=y_h$ is measured from the hadron side and  
$Q^2=Q_l^2$ is measured from the lepton side.
Then the rescaled variables for initial and final state radiation 
are given by
\bea
\label{eqRESC1}
&&\hspace{-0.6cm}{\sf ISR~:}~~~\widehat{y} = \frac{y_h}{z},~\widehat{Q}^2 
           = z Q^2_l,~\widehat{S} = zS,~\widehat{x} = zx_m,
           \nonumber \\
&&~~~~~J^I(z) = 1,~~~~~z_0^I = {\rm max}\left\{y_h,Q_0^2/Q_l^2
\right\}~,  
\eea
\bea
\label{eqRESC2}
&&\hspace{-0.6cm}{\sf FSR~:}~~~\widehat{y} = y_h,~\widehat{Q}^2
           = \frac{ Q^2_l}{z},~\widehat{S} 
           = S,~\widehat{x} = \frac{x_m}{z}~,
           \nonumber \\
& &~~~~~J^F(z) = \frac{1}{z}~,~~~~~z_0^F = x_m~.
\eea 
Here $J^{I,(F)}(z)$ are the Jacobians for initial (final) state 
radiation and $z_0$ denotes the lower bound of the rescaling variable.
$Q_0^2$ is introduced as a cut on $Q_h^2$ to keep the process duly 
deep inelastic, i.e. to avoid significant contributions of the Compton 
peak.
In the subsequent section, we frequently use the following 
shorthand notation for a function with rescaled variables 
in its argument~:
\bea
\widetilde{F}_{I,F}(y,Q^2) 
= F\left(y=\widehat{y}_{I,F},Q^2=\widehat{Q}^2_{I,F}\right) , 
\eea 
where $I,F$ indicate ISR and FSR rescaling. 

\section{NLO corrections}
We parameterize the $k$-th order differential cross section as
\bea
\frac{d^2 \sigma^{(k)}}{dy_h dQ_l^2} \hspace{-1mm} = \hspace{-1mm}
\sum^k_{l=0}\left(\frac{\alpha}{2\pi}\right)^k \hspace{-1mm}
\ln^{k-l} \hspace{-1mm}
\left(\frac{Q^2}{m_e^2}\right)C^{(k,l)}(y,Q^2)~,
\eea
with $C^{(0,0)}(y,Q^2)$ denoting Born cross section.  
$C^{(1,0)}(y,Q^2)$ and $C^{(2,0)}(y,Q^2)$ were calculated in
\cite{JB94}. The $O(\alpha)$ non-logarithmic term 
$C^{(1,1)}(y,Q^2)$ was derived in Ref.~\cite{ABKR}. We re-calculated 
these corrections \cite{JBHK2} and agree with the previous results.

NLO corrections $C^{(1,1)}(y,Q^2)$ are obtained using RG equations 
for mass factorization and charge renormalization. 
This method was first implemented in \cite{BBN} for initial state 
corrections to $e^+e^-$ annihilation, a single differential cross section 
in the $s$--channel. We deal with double--differential distributions
for a $t$--channel process. 
At first the scattering cross section is decomposed as follows~:
\bea
\label{eqCONV}
\frac{d^2 \sigma}{dy_h dQ_l^2}\!\!\! &=& \!\!\!
\frac{d^2 \sigma^0}{dy_h dQ_l^2} \otimes \Biggl\{ 
  \Gamma^I_{ee} \otimes \hat{\sigma}_{ee} \otimes \Gamma^F_{ee}
+ \Gamma^I_{\gamma e} \otimes \hat{\sigma}_{e\gamma} \otimes \Gamma^F_{ee}
+ \Gamma^I_{ee} \otimes \hat{\sigma}_{\gamma e} \otimes 
\Gamma^F_{e \gamma}
\Biggr\} \non\\
\eea
with $\Gamma^{I,F}_{ij}(z,\mu^2/m^2_e)$ the initial and final state
operator matrix elements and $\hat{\sigma}_{kl}(z,Q^2/\mu^2)$ the 
respective Wilson coefficients which obey the representations
\bea
\Gamma^{I,F}_{ij}\left(z,\frac{\mu^2}{m_e^2}\right) &=&
\delta(1-z) + \sum_{m=1}^{\infty} \left (\frac{\alpha}{2\pi} \right)^m
\sum_{n=0}^m
\Gamma_{ij}^{I,F(m,n)}(z)
\ln^{m-n}\left(\frac{\mu^2}{m_e^2}\right) \\
\hat{\sigma}_{kl}\left(z,\frac{Q^2}{\mu^2}\right) &=&
\delta(1-z) + \sum_{m=1}^{\infty} \left(\frac{\alpha}{2\pi} \right)^m
\sum_{n=0}^m 
\widehat{\sigma}_{kl}^{(m,n)}(z)
\ln^{m-n}\left(\frac{Q^2}{\mu^2}\right)~,
\eea
where $j(l)$ denotes the incoming and $i(k)$ the outgoing particle.
In the cross section (\ref{eqCONV}), the $\mu^2$--dependences cancel each
other and the final expression expands in $\alpha/2\pi$ and 
$\ln(Q^2/m^2)$, to $O(\alpha^2 L)$.
There are several contributions to NLO corrections~:
\begin{itemize}
\item[i~] LO initial and  final state radiation off 
$C^{(1,1)}_{ee}(y,Q^2)$
\item[ii~]coupling constant renormalization of $C^{(1,1)}_{ee}(y,Q^2)$
\item[iii~]LO initial state splitting of $P_{\gamma e}$ 
at $C^{(1,1)}_{e \gamma}(y,Q^2)$
\item[iv~]LO final   state splitting of $P_{e \gamma}$ at 
$C^{(1,1)}_{\gamma e}(y,Q^2)$
\item[v~]NLO initial and  final state radiation off 
$C^{(0,0)}_{ee}(y,Q^2)$~.
\end{itemize}
The first contribution $C^{(2,1)}_{\sf i}(y,Q^2)$ is 
\bea
C^{(2,1)}_{\sf i}(y,Q^2)
&=&\int_0^1 dz P_{ee}^0\left[\theta(z-z_0^I)
J^I \widetilde{C}^{(1,1)}_I(y,Q^2)-{C}^{(1,1)}(y,Q^2)\right] \\
&& 
+\int_0^1 dz P_{ee}^0\left[\theta(z-z_0^F)
J^F\widetilde{C}^{(1,1)}_F(y,Q^2)-{C}^{(1,1)}(y,Q^2)\right],\non
\eea
where $P_{ee}^0(z)$ is the LO splitting function:
\bea
P_{ee}^0(z)      = \frac{1+z^2}{1-z}~.
\eea
The QED coupling is renormalized as 
\bea
\label{RUN}
\alpha(\mu^2) = \alpha(m_e^2) \left[1 -\frac{\beta_0}{4\pi} 
\ln\left(\frac{\mu^2}{m_e^2}\right)
\right]~,
\eea
with $\beta_0 = -4/3$ and the second contribution $C_{\sf ii}(y,Q^2)$ 
is given by 
\bea
\label{C2ii}
C^{(2,1)}_{\sf ii}(y,Q^2) =  - \frac{\beta_0}{2} C^{(1,1)}(y,Q^2)~.
\eea
In $C^{(2,1)}_{\sf iii,iv}(y,Q^2)$ there appear new subprocesses~:
\bea
C_{\sf iii}^{(2,1)}(y,Q^2)
=\int_{z_0^I}^1dzP_{\gamma e}^0(z)J^{I}(z)
\widetilde{C}^{(1,1)}_{e\gamma}(y,Q^2)\\
C_{\sf iv}^{(2,1)}(y,Q^2)
=\int_{z_0^F}^1dzP_{e \gamma}^0(z)J^{F}(z)
\widetilde{C}^{(1,1)}_{\gamma e}(y,Q^2) ,
\eea
where $P_{\gamma e}^0$ and $P_{e \gamma}^0$ are LO off-diagonal 
splitting functions
\bea
P_{\gamma e}^0 = \frac{1+(1-z)^2}{z}~,~~~~~~~~
P_{e \gamma}^0 = z^2+(1-z)^2~.
\eea
$C^{(1,1)}_{e \gamma}(y,Q^2)$ and $C^{(1,1)}_{\gamma e}(y,Q^2)$ 
are defined in the same way as $C^{(1,1)}(y,Q^2)$ 
and their explicit expressions are given in \cite{JBHK2}.  \\

The last contribution $C^{(2,1)}_{\sf v}(y,Q^2)$ is given by
\bea
C^{(2,1)}_{\sf v}(y,Q^2) &=&  
\int_0^1 
P_{ee,S}^{1,NS,{\rm OM}}(z) \left[\theta\left(z-z_0^I\right) J^I(z)
\widetilde{C}^{(0,0)}_I(y,Q^2) \right.\nonumber\\ & & \left.
- C^{(0,0)}(y,Q^2)\right] 
+ \int_{z_0^I}^1
P_{ee,S}^{1,PS,{\rm OM}}(z)  J^I(z)
\widetilde{C}^{(0,0)}_I(y,Q^2) \\
&+ &\int_0^1
P_{ee,T}^{1,NS,{\rm OM}}(z) \left[\theta\left(z-z_0^F\right) J^F(z)
\widetilde{C}^{(0,0)}_F(y,Q^2) \right. \nonumber\\ & & \left.
- C^{(0,0)}(y,Q^2)\right] 
+ \int_{z_0^F}^1
P_{ee,T}^{1,PS,{\rm OM}}(z)  J^F(z)
\widetilde{C}^{(0,0)}_F(y,Q^2)~.
\non
\eea
Here $P_{ee,S,T}^{1}(z)$  denote the space-- and time--like NLO QED 
splitting functions of the non--singlet $(NS)$ and pure--singlet $(PS)$ 
channels in the on--mass--shell scheme which are obtained from the 
$\overline{\rm MS}$-scheme~\cite{NLOSP}~ by
\bea
P_{ee,S,T}^{1,NS,{\rm OM}}(z)
&=& P_{ee,S,T}^{1,NS,\overline{\rm MS}}(z) 
+ \frac{\beta_0}{2} \Gamma_{ee}^{S,T,(1,1)}(z)~, \\
\Gamma_{ee}^{S,T,(1,1)}(z) &=& -2 \left[\frac{1+z^2}{1-z}\left(\ln(1-z)
+\frac{1}{2}\right)\right]~,
\eea
and 
$P_{ee,S,T}^{1,PS,{\rm OM}}(z) = P_{ee,S,T}^{1,PS,\overline{\rm MS}}(z)$.
We would like to remark that both the lepton--hadron interference term 
and the pure hadronic QED corrections, although apparently not widely 
known, are small. Already in $O(\alpha)$ their inclusion will only lead to
a marginal change of the present result. In the case of the purely 
hadronic corrections details are explained e.g. in [1a].

\vspace*{-3mm}
\section{Conclusions}
We calculated the $O(\alpha^2L)$ leptonic QED corrections to deep 
inelastic electron-nucleon scattering in the mixed variables. 
With the help of the RGE decomposition, the corrections are expressed  
as the convolutions of the splitting functions with the Born or 
1--loop cross sections. This method generalizes earlier
investigations for ISR in $e^+e^-$ annihilation \cite{BBN} and includes 
both space- and time-like splitting functions and Wilson coefficients.

\vspace*{-3mm}

\end{document}